%Paper: astro-ph/9307012
%From: BEST@sns.ias.edu
%Date: 06 Jul 1993 16:33:53 -0400 (EDT)

%\input phyzzx
\hoffset=0.375in
\overfullrule=0pt

\def\htp{{.\hskip-2pt ''}}
\centerline{\bf Geometry of the HST Planetary Camera Field}
\bigskip
\bigskip
\centerline{Andrew Gould and Brian Yanny}
\medskip
\centerline{Institute For Advanced Study}
\centerline{Princeton, New Jersey 08540}
\bigskip
%\endpage
%\doublespace
\singlespace
\centerline{\bf ABSTRACT}
	We present a solution for the relative positions and orientations
of the four CCD chips on the Hubble Space Telescope (HST) Planetary Camera
(PC).  An accurate solution is required when matching HST images
with ground-based images or with one another.
The solution is accurate to about 1/4 PC pixel or about $0\htp 01$,
a 30-fold improvement over the best previous solution.
The CCDs are rotated relative to one another by up to $1^\circ$.  The
solution is based on images taken between December 1990
and June 1992 and is stable over that entire period.  We find that
the pixel size is $0\htp 04373\pm 0\htp 00004$ based on comparison
with HST Guide Stars, in good agreement with previous Guide-Star
based calibrations but $\sim 1\%$ smaller than a globular-cluster
based calibration.

subject headings: astrometry,  instrumentation: detectors
\endpage

\chapter{Introduction}
\normalspace

	It is possible to measure relative positions on a single CCD chip of
the Hubble Space Telescope (HST) Planetary Camera (PC) with an accuracy of
0.1 pixels or 4 mas (Yanny et al.\ 1993).  Since the PC is a four-CCD
mosaic, it is important to know the orientation and separation of the four
CCDs relative to one another.  The approximate values are given by
Faber (1991), but they are several pixels from the true values.
When matching HST images with ground-based images or with each other,
a more precise transformation is required
(e.g.\ Yanny et al.\ 1993;
Gould et al.\ 1994).

	Here we present a set of transformations from HST PC coordinates
to a single Euclidean coordinate system for the entire PC field.
These will be useful to both current and archival users of HST PC data.
The transformations were obtained by two independent methods.  These
were found to be consistent with each other and were combined to produce
a single best estimate.

	In the first method, we
analyzed 155 ground-based images containing a total of 264 widely
separated star-quasar pairs (including multiple images of the same pair)
that had been
imaged by the HST PC as part of a Snapshot Survey
search for gravitationally lensed quasars  (Bahcall et al.\ 1992;
Maoz et al.\ 1993).
The Snapshot Survey stars were imaged between 1991 April 19 and 1992 June 30.
The
positions of the stars and quasars on the HST images had been measured by
Gould, Bahcall, \& Maoz (1993).  The ground-based images were taken between
1993 March 4 and 8 on the
0.9m telescope with the T2KA CCD at the Kitt Peak National Observatory as
part of an
effort to obtain $V-I$ colors for the Snapshot Survey stars (Gould, Bahcall, \&
Maoz, in preparation).  The scale of the CCD was measured against HST Guide
Star Catalogue stars and found
to be $0\htp 6855/$pixel with an internal error of $0.1\%$ (see below).

	In the second method, we obtained independent
solutions for two HST PC images of M15, one taken on 1990 December 10 and one
on 1991 April 11.  Positions of stars in these fields had been determined
by Yanny et al.\ (1993).  The solutions were found by comparing the positions
of 61 and 79 relatively isolated stars in these images with the positions
in a ground-based image that was obtained by D.\ Schneider on the Palomar
5m telescope in $1''$ seeing at a scale of $0\htp 334/$pixel
in the fall of 1986.

	The two transformations based on
the M15 images agree with each other within errors and were combined
into a single solution by weighting the individual solutions by their
inverse covariance matrices (e.g.\ Gould 1989).
The combined M15 solution agrees with the solution based on the
Snapshot Survey star-quasar pairs within errors.  These two solutions
were then combined into a single solution which is given below.

	We subdivided the HST
exposure data by three-month periods to see if the solution was time-dependent.
Each of the sub-periods agrees with the general solution within errors.
We conclude that the solution is stable
during the entire span of observations from 1990 December 10 to
1992 June 30.

\chapter{The Transformations}

	The four PC CCD chips are numbered 5, 6, 7, and 8, and are arranged
counter-clockwise.  The pixels on each chip form an $800\times 800$ grid with
the position (800,800) in the corner of the field.  See Figure 1.

	We construct a common coordinate system by taking the $x$
and $y$ axes of chip 5 as the axes of the common system.  We label
these $x_5$ and $y_5$ respectively.  We write a general transformation
from pixel numbers $(x_n,y_n)$ on chip $n=6,7,8$ to $x_5$ and $y_5$,
$$ \eqalign{x_5 = & x_n\cos\phi_n - y_n\sin\phi_n + \Delta X_n\cr
y_5 = & x_n\sin\phi_n + y_n\cos\phi_n + \Delta Y_n,}\eqn\xnyn$$
and solve for the three relative rotations $\phi_n$ and six relative
offsets $\Delta X_n$ and $\Delta Y_n$.

We find
$$ \eqalign{\Delta X_6 = & 69.43\pm 0.23\qquad \Delta Y_6 = -0.32
\pm 0.19\qquad \phi_6 = 90.223^\circ\pm 0.027^\circ\cr
\Delta X_7 = & 65.72\pm 0.27\qquad \Delta Y_7 = \ 62.65
\pm 0.28\qquad \phi_7 = 181.032^\circ\pm 0.031^\circ\cr
\Delta X_8 = & \ 6.28\pm 0.28\qquad \Delta Y_8 = \ 50.35
\pm 0.31\qquad \phi_8 = 270.609^\circ\pm 0.037^\circ.}\eqn\tran$$
\FIG\one{Relative positions and orientations of the 4 HST PC CCD chips.
The chips are outlined at different line weightings for clarity.  The
orientations of the $x$ and $y$ axes are shown for each chip individually.
The coordinates on the border of the figure represent the $(x_5,y_5)$
global solution of equation \xnyn.}
Figure \one\ shows the relative orientation of the four CCD chips as
given by equation \xnyn.

	The errors $\sigma_i$ listed in equation \xnyn\ are obtained from
the diagonal elements of
the covariance matrix, $c_{ij}$
$$\sigma_i \equiv (c_{ii})^{1/2}.\eqn\sigi$$
The covariance matrix in
turn is based upon the estimated errors in measuring the stellar
positions.  (See e.g.\ Press et al.\ 1989 for methods of calculating
$c_{ij}$.)\ \ For the Snapshot Survey images, these estimates were found by
adding two sources of error in quadrature.
The first source is the position-fitting error based on the
measured width of the point spread function, gaussian statistics, and
assuming infinite resolution. The second source of error is a uniform term
of cosmic scatter which takes account of the finite resolution
and any other unrecognized sources of error.  The cosmic
scatter was determined by requiring that the $\chi^2$ per degree
of freedom be unity, and was found to be 0.044 ground-based
pixels, or about $0\htp 03$.  For the M15 images all stellar positions
were assigned the same cosmic scatter which was again normalized to
make the $\chi^2$ per degree of freedom equal to unity.  The cosmic scatter
was found to be 0.10 ground-based pixels or about $0\htp 0 3$ for each
image.

\chapter{Correlated Parameters}

	For many purposes, the knowledge that the errors in equation \tran\ are
small, $\sim 1/4$ pixel, will be sufficient.  Sometimes, however, one will
want to obtain accurate individual error estimates.  In these cases, the
full covariance matrix, $c_{ij}$, is required.  We find
%$$\eqalign{&{c_{ij}\over\sqrt{c_{ii}c_{jj}}} =\cr & \left(\matrix{
$$\tilde c_{ij}= \left(\matrix{
  1.00 & -0.06 &  0.66 &  0.61 &  0.03 &  0.39 &  0.35 &  0.00 &  0.32 \cr
 -0.06 &  1.00 & -0.04 & -0.18 &  0.36 & -0.32 & -0.23 &  0.36 & -0.24 \cr
  0.66 & -0.04 &  1.00 &  0.32 & -0.25 &  0.57 &  0.29 & -0.26 &  0.48 \cr
  0.61 & -0.18 &  0.32 &  1.00 & -0.28 & -0.11 &  0.27 &  0.02 &  0.23 \cr
  0.03 &  0.36 & -0.25 & -0.28 &  1.00 &  0.22 & -0.12 &  0.48 & -0.21 \cr
  0.39 & -0.32 &  0.57 & -0.11 &  0.22 &  1.00 &  0.30 & -0.20 &  0.39 \cr
  0.35 & -0.23 &  0.29 &  0.27 & -0.12 &  0.30 &  1.00 &  0.23 & -0.28 \cr
  0.00 &  0.36 & -0.26 &  0.02 &  0.48 & -0.20 &  0.23 &  1.00 & -0.67 \cr
  0.32 & -0.24 &  0.48 &  0.23 & -0.21 &  0.39 & -0.28 & -0.67 &  1.00 \cr}
%\right),}
\right),
\eqn\covmat$$
where $\tilde c_{ij}$ is the matrix of correlation coefficients
$$\tilde c_{ij} \equiv {c_{ij}\over \sigma_i \sigma_j},\eqn\ctilde$$
and where we have ordered the vector of parameters by
$$ a_i = (\Delta X_6,\Delta Y_6,\phi_6,
\Delta X_7,\Delta Y_7,\phi_7,
\Delta X_8,\Delta Y_8,\phi_8).\eqn\adef$$
Because the correlation coefficients are large, one should use the full
covariance matrix when estimating the errors in any measurement.
For example, the variance of the difference in position
in the $x_5$ direction of
two stars with positions $x_5^a$ and $x_5^b$ is
$${\rm var}(\delta x_5)\equiv{\rm var}(x_5^a-x_5^b) = \sum_{i=1}^9\sum_{j=1}^9
{\partial (x_5^a - x_5^b)\over \partial a_i}\,c_{ij}\,
{\partial (x_5^a - x_5^b)\over \partial a_j},\eqn\covari$$
where $x_5^a$ and $x_5^b$ are to be regarded as functions of the $a_i$
through equation \xnyn.  The variance of the distance between two points
is ${\rm var}(\delta r)\equiv{\rm var}[(\delta x_5)^2+(\delta y_5)^2]^{1/2}$,
$$ {\rm var}(\delta r) =
{(\delta x_5)^2{\rm var}(\delta x_5)+(\delta y_5)^2{\rm var}(\delta y_5)
+ 2\delta x_5\delta y_5{\rm covar}(\delta x_5,\delta y_5)\over
(\delta x_5)^2 + (\delta y_5)^2}.\eqn\covarij$$
We find that the errors in $\delta x_5$, $\delta y_5$, and $\delta r$
typically range between 0.18 and 0.37 PC pixels.
If the off-diagonal elements are ignored,
the variances will generally be overestimated, often by a factor
$\sim 2$ or more, because the parameters are strongly correlated.

\chapter{Pixel Size Measurement}

	In addition to the 9 parameters listed above, the Snapshot Survey
solution contains three additional parameters.  These are a scale factor
to transform between the space-based and the ground-based
pixel scales, and a two-parameter (linear) fit to the
relative position angle of the HST and ground-based images.
We discuss these in turn.

	From the general solution we find that the HST PC pixels
are a factor $0.06379\pm 0.00003$ smaller than the ground-based pixels.
We measured the separation in ground-based
pixels of four pairs of Landolt (1991) standards, (100-162, 100-267),
(PG1528+062, PG1528+062A), (PG1528+062, PG1528+062B), and
(PG1528+062A, PG1528+062B), finding 185.63, 43.934, 226.06, and
207.42
pixel separations, respectively.
Comparing these with the angular separations $127\htp 06$,
$30 \htp 084$, $155\htp 21$, and $142 \htp 32$ as determined from the
HST Guide Star Catalog (see Taff et al.\ 1990), we find that the size of an
HST PC pixel is
$$ \rm pixel = 0\htp 04373\pm 0\htp 00004\eqn\hstpix$$
where the error is determined from the scatter among the four
measurements.  This is in good agreement with the determination
of Monet based on the Guide Star Catalog positions, but is
$\sim 1\%$ smaller than another determination of Monet, $0\htp 0442$, based
on measurements of $\omega$ Cen (Faber 1991).

	We note that in deriving equation \xnyn\ from the
Snapshot Survey images, we {\it assumed}
that the pixel size is the same on all four chips.  In principle the
optical properties of the system might cause these scales to differ
on different chips.  In the M15 solutions, however, we allowed the pixel
size to vary as a free parameter and found it to be the same
on all four chips to within $\lsim 0.2\%$.

	Finally, in deriving the solution from the Snapshot Survey
images, we find that the position angle of the HST chip 5 $y$-axis (as
recorded in the data-file header provided by the Space Telescope Science
Institute) does not predict with perfect accuracy
the relative position angles of the stars in the ground-based
images.  Instead, we find a rotation, $\psi$, of chip 5 relative to the
ground-based image
which depends on declination.  Specifically,
$$\psi = -0.418^\circ\pm 0.003^\circ - (0.0031\pm 0.0001)\,\delta,\eqn\psidef$$
where $\delta$ is the declination in degrees.  A zero-point offset of this
magnitude is expected because there is no reason for chip 5 to be better
aligned with the nominal position angle than are the other chips.  However,
the trend with declination is not expected.  It is not due to precession,
for example. We are not able to
tell whether this trend is caused by misalignment of the ground-based
telescope, a flaw in the HST guiding, or a combination of both.  We
report equation \psidef\ for completeness.  It has no impact on the
general solution \xnyn.

{\bf Acknowledgements:}  We would like to thank Buell Jannuzi for
making a number of helpful suggestions and Don Schneider for providing us
with an image of M15.
Work by AG was supported by the National Science Foundation (NSF)
(PHY~92-45317).   BY was supported by Hubble Fellowship
HF-1013.01-90A.  The work is based in part
on observations with the NASA/ESA Hubble Space Telescope, obtained
at the Space Telescope Science Institute, which is operated by the
Association of Universities for Research in Astronomy, Inc. (AURA), under
NASA contract NAS5-26555, and in part by observations at the Kitt
Peak National Observatory operated by AURA under agreement with the
NSF.

\endpage

\Ref\Bsnap{Bahcall, J.\ N., Maoz, D., Doxsey, R., Schneider, D.\ P.,
Bahcall, N.\ A., Lahav, O., \& Yanny, B.\ 1992, ApJ, 387, 56}
\Ref\Faber{Faber, S.\ M., editor 1991, ``Final Orbital/Science Verification
Report by the WF/PC Investigation Definition Team'',
(Baltimore: Space Telescope Science Institute)}
\Ref\Gould{Gould, A.\ 1989, ApJ, 341, 748}
\Ref\GBM{Gould, A., Bahcall, J.\ N., \& Maoz, D.\ 1993, ApJS, 88, in press}
\Ref\GBMYK{Gould, A., Bahcall, J.\ N., Maoz, D., \& Yanny, B.\
1994, ApJ, to be submitted}
\Ref\lan{Landolt, A., U.\ 1992, AJ, 104, 372}
\Ref\Maozc{Maoz, D.,
%Bahcall, J.\ N., Schneider, D.\ P.,
%Bahcall, N.\ A., Djorgovski, R.\ Doxsey, R., Gould, A.,
%Kirhakos, S., Meylan, G., \& Yanny, B.\
et al.\ 1993, ApJ, 409, 28}
\Ref\pr{Press, W.\ H., Flannery, B.\ P., Teukolsky, S.\ A.,
\& Vetterling, W.\ T.\ 1989, Numerical Recipes, pp. 511-512
(Cambridge: Cambridge Univ.\ Press)}
\Ref\ta{Taff, L. G. et al.\
%, Lattanzi, M. G., Bucciarelli, B., Gilmozzi, R., McLean, B. J.,
%Jenkner, H., Laidler, V. G., Lasker, B. M., Shara, M. M., \& Sturch, C. R.
1990 ApJ 353, L45}
\Ref\Yanny{Yanny, B., Guhathakurta, P., Schneider, D.\ P.,
\& Bahcall, J.\ N.\  1993, AJ, submitted}
\refout
\figout
\end